\documentclass[12pt]{article}
\usepackage{amssymb}
\usepackage{cite}
\usepackage{mathrsfs}
\renewcommand{\theequation}{\arabic{section}.\arabic{equation}}

\begin{document}



\def\a{\alpha}
\def\b{\beta}
\def\d{\delta}
\def\e{\epsilon}
\def\g{\gamma}
\def\h{\mathfrak{h}}
\def\k{\kappa}
\def\l{\lambda}
\def\o{\omega}
\def\p{\wp}
\def\r{\rho}
\def\t{\tau}
\def\s{\sigma}
\def\z{\zeta}
\def\x{\xi}
\def\V={{{\bf\rm{V}}}}
 \def\A{{\cal{A}}}
 \def\B{{\cal{B}}}
 \def\C{{\cal{C}}}
 \def\D{{\cal{D}}}
\def\G{\Gamma}
\def\K{{\cal{K}}}
\def\O{\Omega}
\def\R{\bar{R}}
\def\T{{\cal{T}}}
\def\L{\Lambda}
\def\f{E_{\tau,\eta}(sl_2)}
\def\E{E_{\tau,\eta}(sl_n)}
\def\Zb{\mathbb{Z}}
\def\Cb{\mathbb{C}}

\def\R{\overline{R}}

\def\be{\begin{eqnarray}}
\def\ee{\end{eqnarray}}
\def\ba{\begin{array}}
\def\ea{\end{array}}
\def\no{\nonumber}
\def\le{\langle}
\def\re{\rangle}
\def\lt{\left}
\def\rt{\right}

\newtheorem{Theorem}{Theorem}
\newtheorem{Definition}{Definition}
\newtheorem{Proposition}{Proposition}
\newtheorem{Lemma}{Lemma}
\newtheorem{Corollary}{Corollary}
\newcommand{\proof}[1]{{\bf Proof. }
        #1\begin{flushright}$\Box$\end{flushright}}

\baselineskip=20pt

\newfont{\elevenmib}{cmmib10 scaled\magstep1}
\newcommand{\preprint}{
   \begin{flushleft}
   \end{flushleft}\vspace{-1.3cm}
   \begin{flushright}\normalsize
   \end{flushright}}
\newcommand{\Title}[1]{{\baselineskip=26pt
   \begin{center} \Large \bf #1 \\ \ \\ \end{center}}}
\newcommand{\Author}{\begin{center}
   \large \bf
Xin Zhang${}^{a}$,~Junpeng Cao${}^{a,b,c}$,~Wen-Li Yang${}^{d,e,f}\footnote{Corresponding author:
wlyang@nwu.edu.cn}$,\\~Kangjie Shi${}^{d,e}$ and~Yupeng
Wang${}^{a,b,c}\footnote{Corresponding author: yupeng@iphy.ac.cn}$
 \end{center}}
\newcommand{\Address}{\begin{center}

     ${}^a$Beijing National Laboratory for Condensed Matter
          Physics, Institute of Physics, Chinese Academy of Sciences, Beijing
           100190, China\\
     ${}^b$Collaborative Innovation Center of Quantum Matter, Beijing,
     China\\
     ${}^c$School of Physical Sciences, University of Chinese Academy of
Sciences, Beijing, China\\
     ${}^d$Institute of Modern Physics, Northwest University,
     Xian 710069, China \\
     ${}^e$Shaanxi Key Laboratory for Theoretical Physics Frontiers,  Xian 710069, China\\
     ${}^f$Beijing Center for Mathematics and Information Interdisciplinary Sciences, Beijing, 100048,  China
   \end{center}}
\newcommand{\Accepted}[1]{\begin{center}
   {\large \sf #1}\\ \vspace{1mm}{\small \sf Accepted for Publication}
   \end{center}}

\preprint
\thispagestyle{empty}
\bigskip\bigskip\bigskip

\Title{Exact solution of the relativistic quantum Toda chain} \Author

\Address
\vspace{1cm}

\begin{abstract}
The relativistic quantum Toda chain model is studied with the
generalized algebraic Bethe Ansatz method. By employing a set of
local gauge transformations, proper local vacuum states can be
obtained for this model. The exact spectrum and eigenstates of the
model are thus constructed simultaneously.

\vspace{1truecm} \noindent {\it PACS:} 02.30.Ik, 03.65.Vf, 02.10.De.

\noindent {\it Keywords}: Toda chain, Generalized Bethe Ansatz, $T-Q$ relation
\end{abstract}

\newpage

\section{Introduction}

The Toda chain model \cite{Toda1,Toda2} is one of the most fascinating integrable models \cite{Henon,Olshanetsky,Flaschka1}, which
plays an important role in theoretical physics and mathematics such as in Yang-Mills field theory \cite{Marshakov} and group theory \cite{Mikhailov}. In the past several decades, considerable attentions have been paid on both the classical Toda chain \cite{Flaschka1,Flaschka2,Flaschka3} and the quantum Toda chain \cite{Gutzwiller,sutherland,Sklyanin,Sklyanin2,Pasquier,Jimbo} because of its unique property: an infinite dimensional model without $U(1)$ symmetry. This unique property makes it almost impossible to derive its spectrum with the conventional Bethe Ansatz methods such as the coordinate Bethe Ansatz and the algebraic Bethe Ansatz since the absence of an obvious reference state. To overcome this difficulty, several remarkable approaches have then been developed such as Gutzwiller's Ansatz \cite{Gutzwiller} for few-body systems, asymptotic Bethe Ansatz \cite{sutherland} and functional Bethe Ansatz \cite{Sklyanin,Pasquier,Sklyanin2}. An interesting issue is that the relativistic Toda chain model \cite{Ruijsenaars,suris,Kundu2,Pakuliak}, which is tightly related to the Seiberg-Witten theory \cite{Marshakov} and the Calabi-Yau manifold \cite{Kharchev,Smirnov,Gerasimov}, is also integrable. However, its exact solution via Bethe Ansatz has not been derived yet\footnote{The exact quantization condition was studied via WKB expansion in \cite{Hua15, K, Kas16}.}.

In this paper, we study  the relativistic quantum Toda chain \cite{suris,Kundu2} described by the Hamiltonian
\be
H=\sum_{n=1}^N\cos(2\eta \hat{p}_n)+\sum_{n=1}^{N}g^2\cos(\eta\hat{p}_n+\eta\hat{p}_{n+1})\,e^{{x}_{n+1}-{x}_n},\label{def H}
\ee
where $\eta$ and $g$ are two generic coupling constants; ${x}_n$ and $\hat{p}_n$ denote the coordinate and momentum operators on site $n$ which satisfy the commutation relation: $[{x}_n,\hat{p}_m]=i\delta_{m,n}$. Notice that the periodic boundary condition ${x}_{N+1}={x}_1$ and $\hat{p}_{N+1}=\hat{p}_1$ is imposed in (\ref{def H}). The non-relativistic Toda chain Hamiltonian $\cal{H}$ can be readily obtained by taking the limit $\eta\to0$ and $g=i\sqrt{2c}\eta\to0$ with $c$ a constant. In this limit the Hamiltonian (\ref{def H}) has the expansion
\begin{eqnarray}
H=N-2\eta^2\,\cal{H}+\cdots,\no
\end{eqnarray}
and the resulting $\cal{H}$ is exactly the Hamiltonian of the non-relativistic Toda chain \cite{Toda1,Toda2} given by
\begin{eqnarray}
{\cal H}=\sum_{n=1}^{N}\hat{p}_n^2+c\sum_{n=1}^{N}e^{{x}_{n+1}-{x}_n}.
\end{eqnarray}
Here we shall adopt the generalized Bethe Ansatz method used in \cite{Takhtadzhan,cao03} to construct the exact spectrum of Hamiltonian (\ref{def H}) in the T-Q formalism.

The paper is organized as follows.  In Section 2, we briefly review the integrability of the relativistic quantum Toda chain by constructing the associated commuting transfer matrix.  A set of gauge transformations and the corresponding local vacuum states are given in Section 3. Some useful commutation relations among the matrix elements of the gauged monodromy matrix are also presented. The exact spectrum given in terms of a homogeneous T-Q relation and the associated Bethe Ansatz equations (BAEs) and eigenstates of the transfer matrix are derived in Section 4. Concluding remarks are given in Section 5. Some useful formulas
and another set of gauge transformations are shown in Appendix A \& B, respectively.


\section{Integrability}
\label{The system} \setcounter{equation}{0}

Let $\mathbf{V}$ denote the local Hilbert space.
The Lax operator of the relativistic quantum Toda chain $L_n(u)\in$ End($\mathbf{C}^2\otimes \mathbf{V}$) with a generic spectral parameter $u$ is defined as \cite{suris,Kundu2}
\be
L_n(u)=\left(
         \begin{array}{cc}
           e^{u-i\eta\hat{p}_n}-e^{-u+i\eta\hat{p}_n} & -ge^{x_n} \\
           ge^{-x_n} & 0 \\
         \end{array}
       \right),\quad n=1,\cdots,N,\label{def L}
\ee
which satisfies the Yang-Baxter relation
\be
R(u-v)(L_n(u)\otimes I)(I\otimes L_n(v))=(I\otimes L_n(v))(L_n(u)\otimes I)R(u-v),\quad n=1,\cdots,N,\label{YBE1}
\ee
where the $R$-matrix $R(u)\in$ End($\mathbf{C}^2\otimes \mathbf{C}^2$) reads
\be
R(u)=\left(
       \begin{array}{cccc}
         \sinh(u+\eta) & 0 & 0 & 0 \\
         0 & \sinh u & \sinh\eta & 0 \\
         0 & \sinh\eta & \sinh u & 0 \\
         0 & 0 & 0 & \sinh(u+\eta) \\
       \end{array}
     \right).\label{def R}
\ee

The monodromy matrix of the model is constructed as
\be
T(u)=L_N(u)\cdots L_1(u)=\left(
     \begin{array}{cc}
       A(u) & B(u) \\
       C(u) & D(u) \\
     \end{array}
   \right).
\label{def T}
\ee
With the help of Eq.(\ref{YBE1}), we conclude that
the monodromy matrix (\ref{def T}) also satisfies the Yang-Baxter equation
\be
R_{1,2}(u_1-u_2)T_1(u_1)T_{2}(u_2)=T_{2}(u_2)T_1(u_1)R_{1,2}(u_1-u_2).\label{YBE}
\ee
Here and below we adopt the standard notaitons: for any matrix $A\in$ End($\mathbf{V}$),
$A_n$ is an embedding operator in the tensor space $\mathbf{V}\otimes\mathbf{V}\otimes\cdots$,
which acts as $A$ on site $n$ and an identity on the other factor spaces; $R_{i,j}(u)$
is an embedding operator in the tensor space, which acts as identity on the factor space except for the $i$-th and $j$-th ones.

As usual, the transfer matrix $t(u)$ of the present model is given by
\be
t(u)={\rm tr}(T(u))=A(u)+D(u)=\sum_{j=0}^Nt_{N-2j}e^{(N-2j)u}.\no
\ee
Eq.(\ref{YBE}) leads to that the transfer matrix $t(u)$ forms a mutually commuting family \cite{Baxter,Korepin,Wang1},
i.e., $[t(u),t(v)]=0$.
Therefore, $t(u)$ serves as a generating functional of conserved quantities.
The Hamiltonian of the relativistic quantum Toda chain (\ref{def H}) can be obtained through the resulting conserved quantities by \cite{Kundu2}
\be
H=-\frac{1}{2}\left(t_{N-2}t_N^{-1}+t_{2-N}t_{-N}^{-1}\right).\label{def H 1}
\ee

\section{Gauge transformation and local vaccum}
\label{Gauge transformation} \setcounter{equation}{0}

For generic complex $\eta$ with $Re(\eta)>0$, to find a proper vacuum state which allows us to perform the algebraic Bethe Ansatz, we introduce the following gauge matrices
\be
&&M_{k}(u)=\left(
         \begin{array}{cc}
           X_{k}(u), & Y_{k}(u) \\
         \end{array}
       \right)=\left(
            \begin{array}{cc}
              \frac{e^{-u-k\eta}}{\sinh(k \eta)} & e^{-u+k\eta} \\[6pt]
              \frac{1}{\sinh(k \eta)} & 1 \\
            \end{array}
          \right),\no\\[8pt]
&&M_{k}^{-1}(u)=\left(
              \begin{array}{c}
                \overline{Y}_{k}(u) \\[6pt]
                \overline{X}_{k}(u) \\
              \end{array}
            \right)=\frac{e^{u}}{2}\left(
                                                      \begin{array}{cc}
                                                        -1 & e^{-u+k\eta} \\[6pt]
                                                        \frac{1}{\sinh(k\eta)} & -\frac{e^{-u-k\eta}}{\sinh(k\eta)} \\
                                                      \end{array}
                                                    \right),\label{def transformation}
\ee
where $k$ is a free complex parameter.
The matrices $L_n(u)$ and $T(u)$ under the gauge transformations behave as
\be
&&\overline{L}^{(n)}_{j,k}(u)=M_{j}^{-1}(u)L_n(u)M_{k}(u)=\left(
                                                 \begin{array}{cc}
                                                   \overline{A}^{(n)}_{j,k}(u) & \overline{B}^{(n)}_{j,k}(u) \\[6pt]
                                                   \overline{C}^{(n)}_{j,k}(u) & \overline{D}^{(n)}_{j,k}(u) \\
                                                 \end{array}
                                               \right),\quad n=1,\cdots,N,\label{def gauge L}\\[8pt]
&&\overline{T}_{j,k}(u)=M_j^{-1}(u)T(u)M_k(u)=\left(
\begin{array}{cc}
\overline{A}_{j,k}(u) & \overline{B}_{j,k}(u) \\[6pt]
\overline{C}_{j,k}(u) & \overline{D}_{j,k}(u) \\
\end{array}
\right).\label{def gauge T}
\ee

Now we define a local wave function on site $n$ as
\be
|\a;n\rangle=e^{-\frac{1}{2\eta}(x_n-\a\eta)^2+\b_nx_n},\quad n=1,\cdots,N,\label{def local wave}
\ee
with $\alpha$ a free parameter and
\be
\b_n=-n-\frac12-\frac{(2n+1)\ln g+in\pi}{\eta}.
\ee
As we consider the $Re(\eta)> 0$ case, the wave function (\ref{def local wave}) is convergent in the whole real $x_n$ line. When $Re(\eta)< 0$ (the Hamiltonian is in fact an even function of $\eta$), we can use another set of gauge transformation listed in Appendix B to obtain the exact solutions of this model. The $Re(\eta)=0$ case will be discussed in Section 4.
With the help of the definitions (\ref{def gauge L}) and (\ref{def local wave}), we can easily check that
\be
&&\overline{B}^{(n)}_{\a_{n+1},\a_n}(u)|\a;n\rangle=0,\\[4pt]
&&\overline{A}^{(n)}_{\a_{n+1},\a_n}(u)|\a;n\rangle=ge^{\frac \eta 2}e^{u-n\delta\eta}|\a+1;n\rangle,\\[4pt]
&&\overline{D}^{(n)}_{\a_{n+1},\a_n}(u)|\a;n\rangle=ge^{\frac \eta 2}e^{-u+n\delta\eta}|\a-1;n\rangle,
\ee
where
\be
&&\delta=-1-\frac{2\ln g+i\pi}{\eta},\label{Def-delta}\\ [4pt]
&&\a_n=\a+n\delta,\quad n=1,\cdots,N+1.\label{def alpha}
\ee

Let us define the global vacuum state
\be
|\a\rangle=\otimes_{n=1}^N|\a;n\rangle.\label{def vacuum state}
\ee
Acting the elements of the gauge transformed matrix (\ref{def gauge T}) on the state (\ref{def vacuum state}), we readily have
\be
&&\overline{B}_{\a_{N+1},\a_1}(u)|\a\rangle=0,\\[4pt]
&&\overline{A}_{\a_{N+1},\a_1}(u)|\a\rangle=a(u)|\a+1\rangle,\\[4pt]
&&\overline{D}_{\a_{N+1},\a_1}(u)|\a\rangle=d(u)|\a-1\rangle,
\ee
where
\be
a(u)=g^Ne^{\frac{N\eta}{2}}e^{Nu-\frac{N(N+1)\delta\eta}{2}},\qquad
d(u)=g^Ne^{\frac{N\eta}{2}}e^{-Nu+\frac{N(N+1)\delta\eta}{2}}.\label{def a&d}
\ee
Therefore $|\a\rangle$ serves as a reference state in the sense of Ref.\cite{Takhtadzhan}.

In addition, with the help of the Yang-Baxter relation Eq.(\ref{YBE}), we can derive  the  commutation relations
\be
&&\overline C_{m',m}(u_1)\overline C_{m'+1,m-1}(u_2)=\overline C_{m',m}(u_2)\overline C_{m'+1,m-1}(u_1),\label{CC relation}\\[4pt]
&&\overline A_{m',m}(u_1)\overline C_{m'+1,m-1}(u_2)=\frac{\sinh(u_1-u_2+\eta)}{\sinh(u_1-u_2)}
\overline C_{m'+2,m}(u_2)\overline A_{m'+1,m-1}(u_1)\no\\[4pt]
&&\quad-\frac{\sinh\eta\sinh((m'+1)\eta-u_1+u_2)}{\sinh(u_1-u_2)\sinh(m'+1)\eta}\overline C_{m'+2,m}(u_1)\overline A_{m'+1,m-1}(u_2),\label{AC relation}\\[4pt]
&&\overline D_{m',m}(u_1)\overline C_{m'+1,m-1}(u_2)=\frac{\sinh(u_1-u_2-\eta)}{\sinh(u_1-u_2)}
\overline C_{m',m-2}(u_2)\overline D_{m'+1,m-1}(u_1)\no\\[4pt]
&&\quad+\frac{\sinh\eta\sinh((m-1)\eta-u_1+u_2)}{\sinh(u_1-u_2)\sinh(m-1)\eta}\overline C_{m',m-2}(u_1)\overline D_{m'+1,m-1}(u_2),\label{DC relation}
\ee
which together with the global vacuum state (\ref{def vacuum state}) allow us to perform the generalized algebraic Bethe Ansatz \cite{Takhtadzhan,cao03}.
The proof of the above commutation relations is given in Appendix A.



\section{Generalized Bethe Ansatz}
\label{Generalized Bethe Ansatz} \setcounter{equation}{0}
Let us consider first the case of a special sequence of $\eta$ taking values
\begin{eqnarray}
\eta=\frac{i\pi (2q-N)}{N+2M}-\frac{2N\ln g}{N+2M}, {~~}M=0,1,2,\cdots {~~}{\rm and} \quad q\in \mathbb{Z},\label{constraint}
\end{eqnarray}
and $|g|<1$ (for $|g|>1$ case we can use the gauge transformation introduced in Appendix B). We remark that when the parameter $\eta$ takes the above discrete values  the identity: $e^{2M\eta}=e^{N\delta\eta}$ holds, which  allows us to
introduce the Bethe-type state
\be
|u_1,\cdots,u_M;\a\rangle=\left\{\prod_{j=1}^M\overline{C}_{k_{\a}+j,k_{\a}-j}(u_j)\right\}|\a\rangle,\label{def state1}
\ee
where
\be
k_{\a}=\a+\delta+M.
\ee
Acting $\overline{A}_{k_{\a},k_{\a}}(u)$ and $\overline{D}_{k_{\a},k_{\a}}(u)$ on the state $|u_1,\cdots,u_M;\a\rangle$ and using the commutation relations (\ref{CC relation})-(\ref{DC relation}), we have
\be
&&\overline{A}_{k_{\a},k_{\a}}(u)|u_1,\cdots,u_M;\a\rangle
=a(u;\{u_l\})|u_1,\cdots,u_M;\a+1\rangle\no\\[4pt]
&&\quad\quad+\sum_{j=1}^Ma_j(u;k_{\a};\{u_l\})|u_1,\cdots,u_{j-1},u,u_{j+1},\cdots,u_M;\a+1\rangle,\label{function1}\\[4pt]
&&\overline{D}_{k_{\a},k_{\a}}(u)|u_1,\cdots,u_M;\a\rangle
=d(u;\{u_l\})|u_1,\cdots,u_M;\a-1\rangle\no\\[4pt]
&&\quad\quad+\sum_{j=1}^Md_j(u;k_{\a};\{u_l\})|u_1,\cdots,u_{j-1},u,u_{j+1},\cdots,u_M;\a-1\rangle,\label{function2}
\ee
where
\be
a(u;\{u_l\})&=&a(u)
\prod_{j=1}^M\frac{\sinh(u-u_j+\eta)}{\sinh(u-u_j)},\no\\[4pt]
d(u;\{u_l\})&=&d(u)\prod_{j=1}^M\frac{\sinh(u-u_j-\eta)}{\sinh(u-u_j)},\no\\[4pt]
a_j(u;k_{\a};\{u_l\})&=&-a(u_j)\frac{\sinh\eta\sinh((k_{\a}+1)\eta-u+u_j)}{\sinh(u-u_j)\sinh((k_{\a}+1)\eta)}\prod_{l\neq j}^M\frac{\sinh(u_j-u_l+\eta)}{\sinh(u_j-u_l)},\no\\[4pt]
d_j(u;k_{\a};\{u_l\})&=&d(u_j)\frac{\sinh\eta\sinh((k_{\a}-1)\eta-u+u_j)}
{\sinh(u-u_j)\sinh((k_{\a}-1)\eta)}\prod_{l\neq j}^M\frac{\sinh(u_j-u_l-\eta)}{\sinh(u_j+u_l)}.\no
\ee

Assume that an eigenstate of the transfer matrix $t(u)$ takes the following form
\be
|\l_1,\cdots,\l_M;\bar\a\rangle\rangle=\sum_{n\in\mathbb{Z}}e^{i(\bar\a+n)\phi}|\l_1,\cdots,\l_M;\bar\a+n\rangle,\label{def state}
\ee
where $\phi$ is a complex parameter and
$\bar\a\neq -\delta+j+\frac{ik\pi}{\eta}$, $j,k\in\mathbb{Z}$.
The definition of transfer matrix $t(u)$ implies that
\be
t(u)={\rm{tr}}(T(u))={\rm{tr}}(\overline{T}_{k,k}(u))=\overline{A}_{k,k}(u)+\overline{D}_{k,k}(u).
\ee
Applying $t(u)$ on the eigenstate (\ref{def state}) and using the relations (\ref{function1}) and (\ref{function2}), we have
\be
&&t(u)|\l_1,\cdots,\l_M;\bar\a\rangle\rangle=\L(u)|\l_1,\cdots,\l_M;\bar\a\rangle\rangle+\sum_j^M\L_j(u)\left\{\frac{\sinh\eta}{\sinh(u-\l_j)}
\sum_{n\in\mathbb{Z}}e^{i(\bar\a+n)\phi}\right.\no\\[4pt]
&&\quad\times\left.\frac{\sinh(u-\l_j-(k_{\bar \a}+n)\eta)}{\sinh(( k_{\bar\a}+n)\eta)}|\l_1,\cdots,\l_{j-1},u,\l_{j+1},\cdots,\l_M;\bar\a+n\rangle\right\},\label{eigen equation}
\ee
where $k_{\bar\a}=\bar\a+\delta+M$.
The function $\L(u)$ in (\ref{eigen equation}) identifies the eigenvalue term
\be
\L(u)=e^{-i\phi}a(u)\prod_{j=1}^M\frac{\sinh(u-\lambda_j+\eta)}{\sinh(u-\lambda_j)}
+e^{i\phi}d(u)\prod_{j=1}^M\frac{\sinh(u-\lambda_j-\eta)}{\sinh(u-\lambda_j)},\label{BA1}
\ee
and the coefficients $\{\L_j(u)|j=1,\cdots,M\}$ of the unwanted terms read
\be
\L_j(u)=e^{-i\phi}a(\l_j)\prod_{l\neq j}^M\frac{\sinh(\l_j-\l_l+\eta)}{\sinh(\l_j-\l_l)}
-e^{i\phi}d(\l_j)\prod_{l\neq j}^M\frac{\sinh(\l_j-\l_l-\eta)}{\sinh(\l_j-\l_l)}.\label{BA2}
\ee
To ensure the state (\ref{def state}) to be an eigenstate of the transfer matrix, we should put $\L_j(u)=0$ which gives rise to the associated BAEs
\be
e^{2i\phi}\frac{d(\l_j)}{a(\l_j)}=\prod_{l\neq j}^M\frac{\sinh(\lambda_j-\lambda_l+\eta)}{\sinh(\lambda_j-\lambda_l-\eta)},\quad j=1,\cdots,M.\label{BE 1}
\ee

To determine the parameter $\phi$, we should note that the transfer matrix $t(u)$ possesses the asymptotic behavior
\be
\lim_{u\rightarrow\pm\infty}t(u)=t_{\pm N}e^{\pm Nu}+\cdots=
(\pm1)^Ne^{\mp i\eta\sum_n\hat{p}_n}e^{\pm Nu}+\cdots.\no
\ee
The commutation relation $[t(u),t_n]=0$ implies that
\be
\lim_{u\rightarrow\pm\infty}\Lambda(u)=(\pm1)^Ne^{\mp i\eta K}e^{\pm Nu}+\cdots,\no
\ee
where $K\equiv\sum_n p_n$  is the total momentum of the system.
The above asymptotic behavior allows us to fix the parameter $\phi$ as
\be
e^{i\phi}=g^Ne^{-NM\eta}e^{\frac{N\eta}{2}+i\eta K},\quad {\rm or }\quad e^{-i\phi}=(-1)^Ng^Ne^{(N+2)M\eta}e^{\frac{N\eta}{2}-i\eta K},
\label{def phi}
\ee
which is  self-consistent with each other  for $\eta$ values  given in equation (\ref{constraint}). The $T-Q$ relation given by (\ref{BA1}) can be rewritten as
\be
\Lambda(u)&=&(ig)^Ne^{\frac{N\eta}{2}}e^{Nu-i\eta K}\prod_{j=1}^M\frac{Q(u+\eta)}{Q(u)}
+(-ig)^Ne^{\frac{N\eta}{2}}e^{-Nu+i\eta K}\prod_{j=1}^M\frac{Q(u-\eta)}{Q(u)},\label{bethe ansatz 2}\\
Q(u)&=&\prod_{j=1}^M\sinh(u-\lambda_j),\no
\ee
where $K\in \mathbb{R}$ is the total momentum of the system  and the BAEs (\ref{BE 1}) become
\be
e^{-2N\lambda_j+2i\eta K}=(-1)^N\prod_{l\neq j}^M\frac{\sinh(\lambda_j-\lambda_l+\eta)}{\sinh(\lambda_j-\lambda_l-\eta)},
\quad\quad j=1,\cdots,M.\label{BE 2}
\ee
The $T-Q$ relation (\ref{bethe ansatz 2}) is the eigenvalue of the transfer matrix with the corresponding eigenstste $|\l_1,\cdots,\l_M;\bar\a\rangle\rangle$ given by (\ref{def state}) and (\ref{def phi}), provided that the parameters
$\{\l_j|j=1,\cdots,M\}$ satisfy the BAEs (\ref{BE 2}).

With the help of the definition (\ref{def H 1}), we can easily obtain the
eigenvalue of the Hamiltonian in terms of the Bethe roots as
\be
E=(e^{-2\eta}-1)\sum_{j=1}^M\cosh(2\lambda_j).\label{energy}
\ee

Now let us consider the generic imaginary $\eta$ case which together with real $g^2$ defines the physically meaningful relativistic quantum Toda chain.  Keeping $N$ fixed and taking $M, q\to\infty$ but with $q/M\to$ finite, the $\eta$ values in (\ref{constraint}) become dense and tend to generic imaginary values. In such a limit, we conclude that the $Q$-function is an infinite product for $|g|\neq 1$, quite similar to that of the non-relativistic Toda chain obtained by Sklyanin \cite{Sklyanin} due to the fact that the model is infinite dimensional and without $U(1)$ symmetry. The resulting eigenvalues and the BAEs read
\be
\Lambda(u)&=&(ig)^Ne^{\frac{N\eta}{2}}e^{Nu-i\eta K}\prod_{j=1}^\infty\frac{\sinh(u-\lambda_j+\eta)}{\sinh(u-\lambda_j)}\no\\[4pt]
&&+(-ig)^Ne^{\frac{N\eta}{2}}e^{-Nu+i\eta K}\prod_{j=1}^\infty\frac{\sinh(u-\lambda_j-\eta)}{\sinh(u-\lambda_j)},\label{bethe ansatz 3}
\ee
and
\be
e^{-2N\lambda_j+2i\eta K}=(-1)^N\prod_{l\neq j}^\infty\frac{\sinh(\lambda_j-\lambda_l+\eta)}{\sinh(\lambda_j-\lambda_l-\eta)},
\quad\quad j=1,\cdots,M,\label{BE 3}
\ee
with the constraint condition
\be
\lim_{u\to\pm\infty}\prod_{j=1}^\infty\frac{\sinh(u-\lambda_j\pm\eta)}{\sinh(u-\lambda_j)}=(ig)^{-N}e^{-\frac{N\eta}{2}}.
\ee
A very special case is that of $|g|=1$. In this case, the wave function is somehow subtle but the spectrum can be extrapolated from (\ref{bethe ansatz 2})-(\ref{BE 2}) by taking the limit $Re(\eta)\to 0^+$. We remark that with the cyclic representation of the Weyl algebra \cite{Ruijsenaars,Pakuliak} (replacing $e^{i\eta \hat p_n}$ and $e^{x_n}$ in the present paper by $e^{\eta \hat p_n}$ and $e^{ix_n}$), the model at roots of unit is reduced to a special case of the quantum $\tau_2$-model (finite dimensional) and can be solved exactly via the off-diagonal Bethe Ansatz \cite{tau2}.

\section{Conclusion}
By employing a set of gauge transformations (\ref{def transformation}), we have obtained the proper vacuum state (\ref{def vacuum state}) which
allows us to use the generalized algebraic Bethe ansatz \cite{Takhtadzhan,cao03} to construct the eigenvalues (\ref{energy}), the corresponding
eigenstates (\ref{def state}) and the associated BAEs (\ref{BE 2}) of the relativistic quantum Toda chain when the parameter $\eta$ takes some discrete values (\ref{constraint}).
Keeping $N$ fixed and taking $M, q\to\infty$ but with $q/M\to$ finite, the $\eta$ values in (\ref{constraint}) become dense and tend to generic imaginary values.
Thus the resulting results give the exact solutions of  the relativistic quantum Toda chain.


\section*{Acknowledgments}

W.-L. Yang thanks M. Marino for helpful communications. The financial supports from the National Natural Science Foundation
of China (Grant Nos. 11375141, 11374334, 11434013, 11425522), BCMIIS, the National Program for Basic Research of
MOST (Grant No. 2016YFA0300603) and the Strategic Priority Research Program
of the Chinese Academy of Sciences are gratefully acknowledged.


\section*{Appendix A: Commutation relations}
\setcounter{equation}{0}
\renewcommand{\theequation}{A.\arabic{equation}}
From the definition of the $R$-matrix (\ref{def R}) and the gauge transformation (\ref{def transformation}),
we can obtain the following useful commutation relations \cite{cao03,Zhang}
\be
&&R_{1,2}(u_1-u_2)X^1_m(u_1)X^2_{m-1}(u_2)=\sinh(u_1-u_2+\eta)X^2_m(u_2)X^1_{m-1}(u_1),\label{relation RXX}\\[4pt]
&&R_{1,2}(u_1-u_2)Y^1_m(u_1)Y^2_{m+1}(u_2)=\sinh(u_1-u_2+\eta)Y^2_m(u_2)Y^1_{m+1}(u_1),\label{relation RYY}\\[4pt]
&&R_{1,2}(u_1-u_2)X^1_{m-1}(u_1)Y^2_m(u_2)=\sinh(u_1-u_2)Y^2_{m+1}(u_2)X^1_m(u_1)\no\\[4pt]
&&\qquad+\frac{\sinh\eta\sinh(m\eta+u_1-u_2)}{\sinh(m\eta)}X^2_{m-1}(u_2)Y^1_m(u_1),\label{relation RXY}\\[4pt]
&&R_{1,2}(u_1-u_2)Y^1_{m}(u_1)X^2_{m+1}(u_2)=\sinh(u_1-u_2)X^2_{m}(u_2)Y^1_{m-1}(u_1)\no\\[4pt]
&&\qquad+\frac{\sinh\eta\sinh(m\eta-u_1+u_2)}{\sinh(m\eta)}Y^2_{m}(u_2)X^1_{m+1}(u_1),\label{relation RYX}\\[4pt]
&&\overline{X}_{m+1}^1(u_1)\overline{X}_m^2(u_2)R_{1,2}(u_1-u_2)=\sinh(u_1-u_2+\eta)\overline{X}^2_{m+1}(u_2)\overline{X}^1_m(u_1),\label{relation XXR}\\[4pt]
&&\overline{Y}_{m-1}^1(u_1)\overline{Y}_m^2(u_2)R_{1,2}(u_1-u_2)=\sinh(u_1-u_2+\eta)\overline{Y}^2_{m-1}(u_2)\overline{Y}^1_m(u_1),\label{relation YYR}\\[4pt]
&&\overline{X}^1_{m-1}(u_1)\overline{Y}^2_m(u_2)R_{1,2}(u_1-u_2)=\sinh(u_u-u_2)\overline{Y}^2_{m+1}(u_2)\overline{X}^1_m(u_1)\no\\[4pt]
&&\qquad+\frac{\sinh\eta\sinh(m\eta+u_1-u_2)}{\sinh(m\eta)}\overline{X}^2_{m-1}(u_2)\overline{Y}^1_m(u_1),\label{relation XYR}\\[4pt]
&&\overline{Y}^1_m(u_1)\overline{X}^2_{m+1}(u_2)R_{1,2}(u_1-u_2)=\sinh(u_1-u_2)\overline{X}^2_m(u_2)\overline{Y}^1_{m-1}(u_1)\no\\[4pt]
&&\qquad+\frac{\sinh\eta\sinh(m\eta-u_1+u_2)}{\sinh(m\eta)}\overline{Y}^2_m(u_2)\overline{X}^1_{m+1}(u_1).\label{relation YXR}
\ee

Multiplying Eq.(\ref{YBE}) with $\overline{X}^1_{m'+1}(u_1)\overline{X}^2_{m'}(u_2)$ from the left and $X^1_m(u_1)X^2_{m-1}(u_2)$ from the right
and using the relations (\ref{relation RXX}) and (\ref{relation XXR}), we arrive at (\ref{CC relation}).
Multiplying Eq.(\ref{YBE}) with $\overline{Y}^1_{m'+1}(u_1)\overline{X}^2_{m'+2}(u_2)$ from the left and $X^1_m(u_1)X^2_{m-1}(u_2)$ from the right
and using the relations (\ref{relation RXX}) and (\ref{relation YXR}), we arrive at (\ref{AC relation}). Multiplying Eq.(\ref{YBE}) with $\overline{X}^1_{m'+1}(u_1)\overline{X}^2_{m'}(u_2)$ from the left and $X^1_{m-2}(u_1)Y^2_{m-1}(u_2)$ from the right
and using the relations (\ref{relation RXY}) and (\ref{relation XXR}), we arrive at (\ref{DC relation}) by exchanging $u_1$ and $u_2$.


\section*{Appendix B: Another gauge transformation}
\setcounter{equation}{0}
\renewcommand{\theequation}{B.\arabic{equation}}

An alternating set of gauge matrices are
\be
&&\widetilde{M}_k(u)=\left(
                    \begin{array}{cc}
                      \widetilde{X}_k(u), & \widetilde{Y}_k(u) \\
                    \end{array}
                  \right)=\left(
                            \begin{array}{cc}
                              \frac{e^{u-k\eta}}{\sinh(k\eta)} & e^{u+k\eta} \\[4pt]
                              \frac{1}{\sinh(k\eta)} & 1 \\
                            \end{array}
                          \right),\\[8pt]
&&\widetilde{M}^{-1}_k(u)=\left(
                            \begin{array}{c}
                              \widehat{Y}_k(u) \\[4pt]
                              \widehat{X}_k(u) \\
                            \end{array}
                          \right)=\frac{e^{-u}}{2}\left(
                                    \begin{array}{cc}
                                      -1 & e^{u+k\eta} \\[4pt]
                                      \frac{1}{\sinh(k\eta)} & -\frac{e^{u-k\eta}}{\sinh(k\eta)} \\
                                    \end{array}
                                  \right).
\ee
Similarly, we define the following transformed matrices
\be
&&\widetilde{L}^{(n)}_{j,k}(u)=\widetilde M_j^{-1}(u)L_n(u)\widetilde M_k(u),\\[6pt]
&&\widetilde{T}_{j,k}(u)=\widetilde M_j^{-1}(u)T(u)\widetilde M_k(u)=\left(
                                                                       \begin{array}{cc}
                                                                         \widetilde{A}_{j,k}(u) & \widetilde{B}_{j,k}(u) \\[6pt]
                                                                         \widetilde{C}_{j,k}(u) & \widetilde{D}_{j,k}(u) \\
                                                                       \end{array}
                                                                     \right).
\ee
Following the similar procedures introduced in Section 3 and Section 4, we can construct the exact solutions for the case of $Re(\eta)< 0$.

\end{document}